\documentclass[aps,twocolumn,showpacs,amsmath,amssymb,pre,superscriptaddress, floatfix]{revtex4}

\usepackage{graphicx}
\usepackage{dcolumn}
\usepackage{bm}
\usepackage{amssymb}
\usepackage{multirow}
\usepackage{color}

\newcommand{\1}{\begin{equation}}
\newcommand{\2}{\end{equation}}
\newcommand{\ea}{\begin{eqnarray}} 
\newcommand{\ee}{\end{eqnarray}}
\newcommand{\4}[2]{{\frac{#1}{#2}}}
 
\newcommand{\Sum}[2]{{\sum\limits_{#1}^{#2}}}

\newcommand{\I}{{ {\rm i}  }}

\begin{document}

\title{
Phoretic Interactions
Generically Induce Dynamic Clusters and Wave Patterns in Active Colloids}

\date{\today}

\author{Benno Liebchen}
\email[]{Benno.Liebchen@ed.ac.uk}\email[]{bliebche@gmail.com}\affiliation{SUPA, School of Physics and Astronomy, University of Edinburgh, Edinburgh EH9 3FD, United Kingdom}
\author{Davide Marenduzzo}\affiliation{SUPA, School of Physics and Astronomy, University of Edinburgh, Edinburgh EH9 3FD, United Kingdom}
\author{Michael E. Cates}\affiliation{DAMTP, Centre for Mathematical Sciences, University of Cambridge, Cambridge CB3 0WA, United Kingdom}

\begin{abstract}
We introduce a representative minimal model for phoretically interacting active colloids.
Combining kinetic theory, linear stability analyses, and a 
general relation between self-propulsion and phoretic interactions in auto-diffusiophoretic and auto-thermophoretic Janus colloids collapses 
the parameter space to two dimensions: area fraction and P\'eclet number. This collapse arises when the lifetime of the self-generated phoretic fields is not too short, and
leads to a 
universal phase diagram showing that 
phoretic interactions {\it generically} induce 
pattern formation in typical Janus colloids, even at very low density. 
The resulting patterns include waves and dynamic aggregates closely resembling the living clusters found in experiments on dilute suspension of Janus colloids. 
\end{abstract}

\maketitle

Chemical signalling between cells is at the heart of many of the remarkable self-organization and pattern formation processes observed in the biological world.
Microorganisms such as {\it Dictyostelium}, which excrete chemicals to which they respond themselves, provide an illustrative example of signalling-driven pattern formation. 
If they swim towards the signalling molecule (chemoattraction), any local accumulation of microorganisms causes an enhanced signal production, in turn recruiting further cells. This creates a positive feedback loop destabilizing the uniform phase (the Keller-Segel instability~\cite{Keller1970,Keller1971}) and leading to the formation of clusters which coarsen indefinitely. 
We recently found that a chemorepulsive response, where microorganisms swim away from the chemical they produce, provides an equally viable, if less intuitive, route to structure formation, resulting in clusters of self-limiting size, moving states and travelling waves~\cite{Liebchen2015}.

A fascinating analogue to biological signalling is provided by the collective behaviour of synthetic autophoretic microswimmers. Such swimmers, often fabricated as Janus colloids, catalyse a chemical reaction on part of their surface and move in the resulting self-produced gradient by diffusiophoresis, or a similar mechanism. 
The resulting gradients then also act on other active particles, inducing chemically mediated (cross-phoretic) many-body interactions. By now, there are several models establishing the analogy between biological and synthetic signalling also formally \cite{Meyer2014, Saha2014, Pohl2014, Liebchen2015, Liebchen2016}.

\begin{figure*}
\begin{center}
\includegraphics[width=0.95\textwidth]{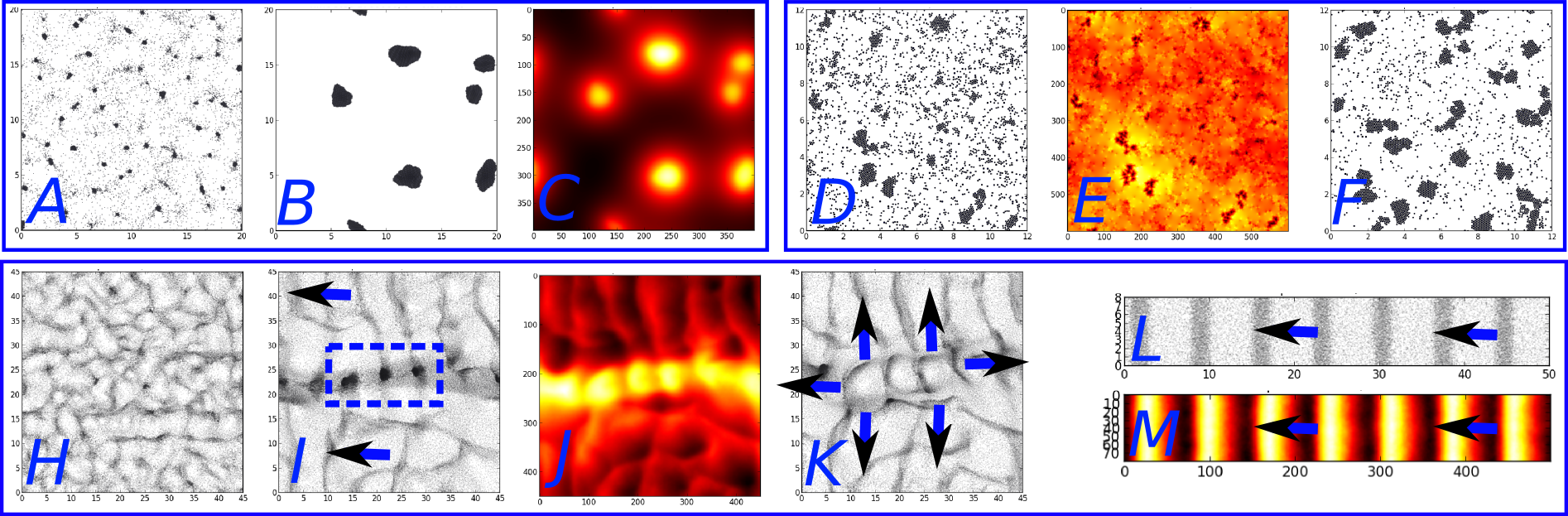}
\caption{\small Generic patterns in phoretic Janus colloids: snapshots from particle-based simulations (movies and parameters in \cite{SM}).
(A-C, movie 1): Attractive phoretic interactions induce clusters at early times (A) which merge (B), accompanied by a colocated phoretic field (C). 
(D-F, movie 5): Dynamic clusters induced by repulsive phoretic interactions (D) surrounded by shells of large phoretic fields (E) which do not coarsen beyond a characteristic size (F).
(H-M, movies 2-4): Colloidal waves (H) pursued by 
self-produced phoretic field waves caging the colloids in dense clusters (I); these clusters act as 
enhanced phoretic producers leading to phoretic clusters (J) which drive
colloids away, and induce escape waves (K). At late times, these wave patterns may settle into regular moving bands of colloids closely followed by phoretic waves (L,M).}
\label{patterns}
\end{center}
\end{figure*}

One notable advantage of synthetic signalling swimmers over their biological counterparts is their conceptual simplicity and controllable design which should render parameter tuning simpler, offering new perspectives for active self-assembly (Fig.~\ref{patterns}). 
Another advantage is that signalling in synthetic swimmers is not restricted to chemical interactions: thermophoretic Janus colloids \cite{Jiang2010}, for example, act as local heat sources and interact via self-produced temperature gradients. Despite this, as we shall see, they can be described by the same equations and allow access to different pattern forming regimes, not accessible for chemical signallers. 

While it is widely believed that phoretic interactions between microswimmers can qualitatively lead to interesting collective behaviour~\cite{Saha2014,Liebchen2015}, little is known about the strength and relevance of these interactions in practical examples of autophoretic colloidal suspensions. 
Most notably, there are now several theoretically plausible mechanisms, based on phoretic concepts, driving the formation of the celebrated, yet mysterious, dynamic ``living'' clusters observed in low density suspensions of active colloids~\cite{Theurkauff2012,Palacci2013,Buttinoni2013,Bialke2015}. 
However, the lack of knowledge of the magnitude or even sign (attractive or repulsive) of the colloidal ``chemotactic'' (or thermotactic) coefficient makes it difficult to understand whether or not cross-phoretic interactions really induce the underlying instability. 

To clarify this situation, the present work addresses the question: `are chemotactic instabilities really there for generic autophoretic colloids'?
To address this, we introduce a representative minimal model for such colloids, the `phoretic Brownian particle' (PBP) model. The PBP model captures the impact of phoretic interactions on the orientations of other colloids, disregards additional but negligible drift effects~\cite{Saha2014,Pohl2014}, and only requires one effective field rather than separate fuel and product fields.

Combining kinetic theory and linear stability analysis, we formulate generic instability criteria for Janus colloids whose phoretic interactions are either attractive (Keller-Segel instability~\cite{Keller1970,Keller1971}) or repulsive (Janus and delay-induced instabilities~\cite{Liebchen2015}). These criteria involve the microscopic parameters of the underlying Langevin equations, and are robust against short-range repulsions, as our large-scale particle-based simulations confirm.

As a key result, we unveil a general relation between self-propulsion and phoretic interactions in self-diffusiophoretic and self-thermophoretic Janus swimmers. This collapses the parameter space, and shows that {\em both} attractive and repulsive phoretic interactions {\em generically} induce structure formation in Janus colloids. Both are therefore generically important for the collective behaviour of Janus colloids.
In contrast to motility-induced phase separation in active Brownian particles (ABPs) \cite{Cates2015} the phoretic patterns we discuss can occur at very low density providing an appealing mechanism to explain the onset of living clusters.

We consider $N$ point-like colloids in quasi-2D, moving with constant speed $v$ (due to self-propulsion) along directions ${\bf p}_i=\left(\cos\theta_i,\sin\theta_i\right); i=1,\ldots,N$. 
These swimming directions change due to rotational Brownian noise and coupling to a phoretic field $c$ which is generated by all other colloids.
This phoretic field is the one relevant for self-propulsion -- a chemical or temperature field for diffusiophoretic or thermophoretic swimmers respectively. (It may involve a combination of fuel and reaction products.)
We define the PBP model by the equations of motion:
\ea
\dot {\bf r}_i(t)&=& v {\bf p}_i \label{rdyn} \\ 
\dot \theta_i(t) &=& \beta {\bf p}_i \times \nabla c +\sqrt{2D_r}\eta_i(t) \label{thdyn}.
\ee
Here, $D_r$ is the rotational diffusion constant and $\eta$ represents unit-variance Gaussian white noise with zero mean; ${\bf a} \times {\bf b} \equiv a_1 b_2 - a_2 b_1$. The phoretic field $c$ is produced at rate $k_0$ by each colloid, and $\beta$ quantifies the coupling to this phoretic field.
When $\beta>0$, particles turn towards the phoretic gradient produced by other colloids, modelling chemoattraction in diffusiophoretic colloids, whereas when $\beta<0$ they swim down the gradient (chemorepulsion).
Generally, Janus swimmers also drift in the phoretic fields produced by other colloids in the same way they drift in their self-produced 
fields \cite{Saha2014,Pohl2014}. However such drifts have little impact on the phase diagram, as we discuss below; hence we neglect them here. 

The phoretic field $c$ evolves as
\ea
\dot c({\bf r},t) &=& D_c \nabla^2 c - k_d c \nonumber \\ 
&+& \Sum{i=1}{N}
\oint
{\rm d}{\bf x}_i \delta({\bf r}-{\bf r}_i(t) - R_0 {\bf x}_i) \sigma({\bf x}_i) \label{ceq}.
\ee
Here, the integral is over the surface of the Janus colloid with radius $R_0$, where $\sigma({\bf x}_i)=k_0/(2\pi R_0^2)$ on the catalytic (coated) hemisphere and zero elsewhere~\footnote{We describe colloids as mechanically point-like, but account for finite particle size effects when describing their phoretic production.}.
We introduced the diffusion constant of the phoretic field $D_c$ and allow for its decay with rate $k_d$. This decay can in effect give screening effects, which we expect to be small when chemical decay is slow.
To reduce the parameter space, we choose time and space units as $t_u=1/D_r$ and $x_u=R_0$, leaving us with five dimensionless numbers alongside the particle density $\rho_0$: (i) the P\'eclet number ${\rm Pe}=v_0/(R_0 D_r)$, measuring the ballistic run length of a colloid in units of its radius; (ii) $B=\beta/(D_r R_0^4)$ comparing the phoretically-induced rotation frequency in (an orthogonal) chemical/thermal unit gradient with rotational diffusion; (iii,iv) $K_0=k_0/D_r; K_d=k_d/D_r$, comparing production and decay rates of the phoretic field to $D_r$, and (v) $\mathcal{D}=D_c/(R_0^2 D_r)$, which measures the timescale that the phoretic field needs to diffuse over the radius of a colloid in units of the inverse rotational diffusion.

\begin{figure*}
\begin{center}
\includegraphics[width=0.9\textwidth]{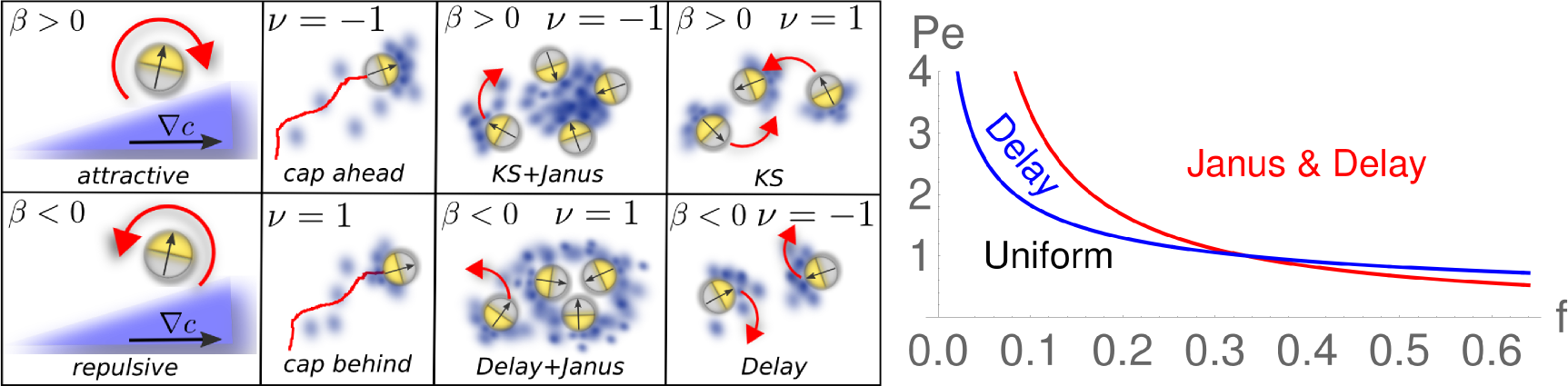} 
\caption{First set of 4 panels (from left): classification of autophoretic Janus colloids in terms of their response to phoretic gradients (attractive/repulsive) and of their swimming direction (cap ahead/behind).
Second set of 4 panels: sketch of the response of Janus colloids to the fields produced by other colloids with indications of which scenarios lead to the sign coefficients required for the Keller-Segel (KS), Janus (Janus) and delay-induced (Delay) instability. 
Right figure: universal phase diagram for quasi-2D repulsively interacting autophoretic colloids depending only on the P\'eclet number (Pe) and the area fraction (f). 
}
\label{class}
\end{center}
\end{figure*}

To understand the collective behavior of autophoretic colloids, we next systematically derive a continuum theory \cite{SM}. The latter generates mean-field equations of motion for the particle density $\rho({\bf x},t)$ and the associated polarization density ${\bf w}({\bf x},t)$,
\ea
\dot \rho &=& - {\rm Pe} \nabla \cdot {\bf w} 
\nonumber \\
\dot {\bf w} &=& -{\bf w} + \4{B\rho}{2}\nabla c - \4{{\rm Pe}}{2}\nabla \rho  + \4{{\rm Pe}^2}{16}\nabla^2 {\bf w} - \4{B^2 |\nabla c|^2}{8} {\bf w} \nonumber \\
&+&
\4{{\rm Pe} B}{16} \left(3 (\nabla {\bf w})^{\rm T}\cdot \nabla c - (\nabla c \cdot \nabla){\bf w} - 3 (\nabla \cdot {\bf w})\nabla c \right)
\nonumber\\ 
\dot c &=& \mathcal{D} \nabla^2 c  + K_0 \rho + \nu \4{K_0}{2}\nabla \cdot {\bf w} - K_d c \label{kineqs}. 
\ee
Here $\nu=1$ for swimmers which move cap-behind, and $\nu=-1$ for cap-ahead swimmers (Fig.~\ref{class}). This model qualitatively resembles the phenomenological model previously considered in~\cite{Liebchen2015}; crucially, however, it provides a microscopic theory here linking all coefficients to microscopic quantities. It also features additional nonlinear terms, which do not affect the linear stability of the uniform phase and the corresponding nonequilibrium phase diagram, but {\em do} influence the emerging patterns. 

Following~\cite{Liebchen2015}, we expect different structure formation scenarios for colloids with attractive and repulsive phoretic interactions, which we now explore through linear stability analysis of the uniform solution $(\rho,{\bf w},c)=(\rho_0,{\bf 0},K_0 \rho/K_d)$ of (\ref{kineqs}).
Specifically, attractive phoretic interactions induce the Keller-Segel (KS) instability described in the introduction 
if $\4{B K_0 \rho_0}{{\rm Pe} K_d}>1 \label{KS}$ \cite{SM}. 
Hence, strong coupling to the chemical field, fast production and high particle density all support this instability. While the KS instability is well-established for microorganisms, our microscopic derivation shows that it also applies to the phoretic interactions relevant to active colloids.

Conversely, for $B<0$ colloids effectively migrate away from each other, and the route to pattern formation depends on the type of phoretic production at their surface. If this is anisotropic, as for Janus colloids, there is an instability inducing clusters of self-limiting size (see~\cite{Liebchen2015} for a discussion of the mechanism). Our linear stability analysis~\cite{SM} shows that this `Janus instability' ensues if
\1 
\4{-B s K_0 \rho_0 (16\mathcal{D}+{\rm Pe}^2)}{\left(4\sqrt{2}\mathcal{D}+{\rm Pe}^2 \right)^2} > 1.
\label{janus}
\2
Remarkably, besides patterns emerging from the Janus instability, a delay in the response of the colloidal swimming direction can trigger a cyclic feedback loop resulting in wave formation~\cite{Liebchen2015}, which is effective if~\cite{SM}
\1
\4{-B \rho_0 K_0 {\rm Pe}}{2\mathcal{D}}>1.
\label{delay}
\2  

To gauge the significance of our instability mechanisms for real Janus colloids, we now reduce the parameter space further. 
For typical diffusiophoretic (thermophoretic) Janus swimmers the reduced diffusion constant is $\mathcal{D} \sim 10^4-10^6 \gg 1$ ($10^6-10^8$), 
suggesting that the $\mathcal{D}\rightarrow \infty$ limit in (\ref{delay}) is physically relevant. However, naively taking this limit would rule out phoretic instabilities altogether.
To see why that approach is invalid, we now consider a test particle, exposed to the phoretic field $c$ produced by other colloids. The gradient of $c$ drives a surface slip velocity on our test particle ${\bf v}_s({\bf r}_s)=\mu({\bf r}) \nabla_{||} c({\bf r})$ (in physical units) 
causing its rotation with a frequency
${\bf \Omega} = \4{3}{2R_0}\langle {\bf v}_s({\bf r}_s) \times {\bf n} \rangle$~\cite{Anderson1989}.
Here, $\nabla_{||}c({\bf r})\equiv (\mathbb{I}-{\bf n}{\bf n})\cdot \nabla c({\bf r})$ is the projection of $\nabla c$ onto the plane tangent to the colloid with unit normal ${\bf n}({\bf r})$, while brackets denote the 
surface average on the test colloid. 
Performing this integral and assuming that $\nabla c$ is constant on the scale of the colloid yields ${\bf \Omega} = \4{3 \nu}{8 R_0}\left(\mu_C - \mu_N \right) {\bf e}\times \nabla c \label{om}$.
Here, $\mu_C,\mu_N$ are the phoretic surface mobilities on the catalytic and the neutral hemisphere of the test colloid, and ${\bf e}$ is the unit vector along its swimming direction. 
Comparing our expression for ${\bf \Omega}$ with (\ref{thdyn}) now yields $\beta=3\nu (\mu_C-\mu_N)/(8R_0)$.

To eliminate from $\beta$ the (usually unknown) mobility coefficients, we now calculate the phoretic field produced by each colloid. We solve the Laplace equation $D_c\nabla^2 c = 0$ with boundary conditions $-D_c {\bf n}\cdot \nabla c = \alpha,0$ 
on the catalytic and neutral caps respectively, and $c(r \rightarrow \infty)=c_0$. This yields in far field $c(r) = c_0 + \4{\alpha R_0^2}{2 D_c r} +\mathcal{O}\left(R_0^3/(D_c r^2)\right)$.
Besides acting on other colloids, this field also drives a (quasi-)slip velocity over the test colloid's own surface leading to self-propulsion with 
${\bf v}=-\langle {\bf v}_s({\bf r}_s)\rangle = -\langle \mu({\bf r}) \nabla_{||} c({\bf r}) \rangle \Rightarrow v =|{\bf v}|= \nu \alpha (\mu_C+\mu_N)/(8 D_c)$ and $\nu ={\rm sign}[(\mu_N+\mu_c)\alpha] $ \cite{Anderson1989,Golestanian2007}.
Combining the former with our previous expression for $\beta$ gives $\beta=3\mu_r D_c v/(R_0 \alpha)$ with the reduced surface mobility $\mu_r=(\mu_C-\mu_N)/(\mu_C+\mu_N)$.
Finally, we compare our expression for $c(r)$ with the steady-state solution of (\ref{ceq}) (screened Poisson equation) for $N=1$,
$c=c_0 + \4{k_0}{4\pi D_c}\4{{\rm exp}[-\sqrt{k_d/D_c}r]}{r}$. This gives $k_0\gtrsim 2\pi R_0^2 \alpha $ with equality for $k_d=0$. Ultimately, 
using $|\mu_r| \approx 1$ for typical Janus colloids at $k_d=0$, we find
\1 \beta = 6\pi R_0 \mu_r \4{v D_c}{k_0} \approx 6\pi R_0 s \4{v D_c}{k_0} \Rightarrow B \approx \4{6 \pi s {\rm Pe} \mathcal{D}}{K_0}\label{betares}\2
whereas $k_d>0$ leads to larger $\beta$ values.

Our key result (\ref{betares}) has notable consequences. 
(i) For typical laser-heated thermophoretic swimmers, $\mu_C\approx 0$~\cite{Bickel2013}, hence $\beta<0$: such swimmers are (thermo)repulsive and therefore a candidate to observe in the laboratory the patterns predicted phenomenologically in \cite{Liebchen2015} for repulsive phoresis. 
(ii) The parameter $\beta$ linearly depends on $D_c$. Thus, the naive approach of taking the $D_c\rightarrow \infty$ limit while keeping $\beta$ constant is inconsistent; phoretic patterns should remain observable even in the limit of fast diffusion. 
(iii) Crucially, Eq.~(\ref{betares}) allows us to eliminate $\beta$ from our instability criteria. Combining Eqs.~(\ref{betares}) and (\ref{delay}), introducing the quasi-2D area fraction $f=\pi R_0^2\rho_0=\pi \rho_0$ (in our units), and performing the limit $\mathcal{D} \rightarrow \infty$
\footnote{Taking this limit is for convenience only; large P\'eclet numbers ${\rm Pe}^2 > \mathcal{D}$ further support instability}
reduces the Janus and the delay-induced instability to:
\1 
-3\nu s{\rm Pe} f > 1; \quad \text{and} \quad -{3} s {\rm Pe}^2 f > 1 \label{JD}
\2

Modulo sign coefficients $s$ and $\nu={\rm sign}[(\mu_N+\mu_C)\alpha]$ (so $\nu=1$ for cap-behind swimmers), these instability criteria depend only on ${\rm Pe}$ and $f$. This massive parameter space collapse yields a universal phase diagram (Fig.~\ref{class}) in which autophoretic colloids, with typical ${\rm Pe} \sim 20-200$ \cite{Theurkauff2012,Palacci2013,Buttinoni2013}, generically form patterns, even at low area fractions of $f\sim 0.01$. Analogously, Eq.~(\ref{betares}) reduces the KS instability for attractive autophoretic colloids to
\1 
{6 \mathcal{D} f}> K_d \label{KSf}
\2
This criterion is fulfilled for $\mathcal{D}\gg 1$ and $K_d\ll 1$ (See SM \cite{SM} for $K_d >1$). Hence, for both self-diffusiophoretic and self-thermophoretic Janus colloids, cross-phoretic interactions generically destabilize the uniform phase. This suggests that models based on purely local interactions such as the ABP model \cite{Romanczuk2012,Cates2015}, are insufficient to capture the collective behaviour of autophoretic systems; indeed they predict onset of structure formation at much higher area fractions than found experimentally~\cite{Theurkauff2012,Palacci2013,Buttinoni2013}. 

Note that the PBP model describes only cross-phoretic {\em alignment} interactions and neglects cross-phoretic {\em drifts}. The latter are a separate source of long-range interactions, and lead to order-$f^3$-corrections of our instability criteria, Eq.~(\ref{JD}). They are negligible at low densities because colloids drift much more slowly in the $1/r^3$-decaying phoretic gradients produced by other colloids than in their self-produced gradients.
 
To test our key findings and to explore their robustness against short-ranged repulsions, we have solved Eqs.~(\ref{rdyn}-\ref{ceq}) numerically (see~\cite{SM} for details). 
Attractive phoretic colloids undergo the KS instability and form small clusters at short times (Fig.~\ref{patterns}A and movie 1), which merge (B) and produce colocated phoretic clusters (C); these coarsen, eventually leaving a single cluster at steady state (not shown). 
This scenario applies even for area fractions $f \ll 0.01$ and is insensitive to parameter variations.
In contrast, repulsive phoretic interactions create a plethora of structures. 
In most cases, the delay-induced instability masks the Janus instability and creates continuously moving patterns. These involve colloidal waves pursued by self-produced phoretic waves; waves often morph into clusters and back to waves. 

At ``early" times, which can last several hours 
in large systems, the delay-induced instability creates waves moving along randomly chosen directions (Fig.~\ref{patterns} H) which coarsen to a characteristic scale (I, movies 2-4).
When these waves collide frontally (movies 2,3), the pursuing phoretic waves act as a cage for (repulsive) particles, compressing them into dense clusters (I, blue rectangle). The high particle density within the cluster enhances the phoretic production, leading, a short while after, to co-located phoretic clusters (J). The high phoretic field $c$ then expels the colloids, so that the clusters explode, initiating a new set of colloidal ring-waves leaving low density regions at the locations of the former clusters (K). 
These waves continue to colloide dynamically and generate a rich pattern of exploding and travelling clusters (the latter emerge from less frontal collisions, movie 3) and of waves which may spontaneously change their direction of motion.
At late times, this type of motion can settle down into a regular pattern of moving bands which are closely pursued by self-produced chemical or heat bands; this scenario can be best observed in simulation boxes of large aspect ratio (L,M; movie 4). 

Finally, we consider a variant of the PBP model in which colloids produce a phoretic field on one hemisphere and consume it with the same rate on the other. This yields $c\propto 1/r^2$ in far field, but leaves (\ref{betares}) unchanged modulo order-one prefactors.
Such zero-net-production (ZNP) colloids might mimic self-electrophoretic swimmers in which a current flows between hemispheres to give a $1/r^2$ far field (see \cite{Brown2016}). They might also model Janus particles whose self-diffusiophoretic motion hinges on a nonlinear threshold effect as might arise in the laser-induced local demixing of a binary fluid near the cap \cite{Buttinoni2012,Buttinoni2013}. Interestingly, for ZNP swimmers the Janus instability, whose linear instability criterion is similar to that derived previously, is no longer pre-empted by the delay-induced instability (Fig.\ref{class}) which requires net phoretic production. 
Our simulations (Fig.~\ref{patterns} D-F and movie 5) 
yield dynamic clusters which are surrounded by self-produced phoretic shells (E) and do not coarsen beyond a certain scale, but continuously emerge and disrupt without ever settling into a steady state. 
This phenomenology resembles the living clusters observed in~\cite{Theurkauff2012,Palacci2013}.

In conclusion, our results show that both attractive and repulsive cross-phoretic interactions {\em generically} induce structure formation in self-diffusiophoretic and self-thermophoretic Janus colloids, even at very low densities. 
This relies on a collapse of parameter space, applying when the lifetimes of the phoretic chemicals are not too small compared to the persistence time of a swimmer. 

While we expect that additional factors like hydrodynamic interactions~\cite{Scagliarini2016,Wagner2017} or fuel depletion will not change our finding that phoretic interactions destabilize the uniform phase at very low volume fractions, they will change the form of the emerging patterns and in particular their length scales. 
Hence further studies will be needed in to clarify the relation between our phoretic patterns and living clusters. 

\paragraph{Acknowledgements}
B.L. gratefully acknowledges funding by a Marie Curie Intra European Fellowship (G.A. no 654908) within Horizon 2020. 
We acknowledge support from EPRSC (grant EP/J007404/1) and thank Aidan Brown for useful discussions on self-electrophoresis.

\pagebreak
\newpage
\onecolumngrid

\begin{center}
{\large \bf Supplementary Material}
\end{center}

\subsection{Kinetic Theory}
Here, we develop a continuum theory for self-propelled Janus colloids, including phoretic interactions  \cite{Liebchen2016sm}. 
We begin with equations (1,2) of the main text in dimensionless form
\ea
\dot {\bf r}_i &=& {\rm Pe}\;{\bf p}_i \label{rdynSM}\\
\dot \theta_i &=& B{\bf p}_i \times \nabla c + \sqrt{2}\xi_i(t) \label{thdynSM}
\ee
which are coupled to a chemical field $c({\bf
 x},t)$ evolving according to (3) in the main text. 

Using It\^{o}s Lemma and following \cite{Dean1996} we derive a continuum equation of motion for the combined $N$-particle probability density $f({\bf r},\theta,t)=\Sum{i=1}{N}\delta({\bf r}-{\bf r}_i(t))\delta(\theta-\theta_i(t))$:
\1
\dot f = - {\rm Pe}{\bf p}\cdot \nabla f +  \partial^2_\theta f - \4{b |\nabla c|}{2} \partial_\theta \left[f \sin(\theta + \delta)\right] - \partial_\theta \sqrt{2f} \eta \label{ffp}
\2
Here $\eta= \eta({\bf r},\theta,t)$ is a unit-variance Gaussian white noise field with zero mean and $\delta={\rm arg}(\partial_y-\I \partial_x)$ (i.e. $\cos(\delta)=-\partial_x c/|\nabla c|;\;\; \sin(\delta)=\partial_y c/|\nabla c|$).
We are mainly interested in mean-field phenomena here and therefore neglect the multiplicative noise term $-\partial_\theta \sqrt{2f}\eta$. Transforming (\ref{ffp}) to Fourier space, yields an equation of motion for the Fourier modes $f_k({\bf r},t)=\int f({\bf r},\theta,t){\rm e}^{\I k \theta} {\rm d} \theta$ of $f$:
\1
\dot f_k({\bf r},t) =-\4{{\rm Pe}}{2} \left[\partial_x \left(f_{k+1}+f_{k-1}\right) - \I \partial_y \left(f_{k+1}-f_{k-1}\right)\right]- k^2 f_k + 
\4{b |\nabla c| k}{2} \left(f_{k+1}{\rm e}^{\I \delta} - f_{k-1} {\rm e}^{-\I \delta} \right) \label{ffpk}
\2

Evaluating (\ref{ffpk}) for $k=0,1,\ldots$ leads to a hierarchy of equations for $\{f_k\}$ with $f_0({\bf x},t)=\rho({\bf x},t)=\int f({\bf x},\theta,t){\rm d} \theta$ 
being the orientation-independent probability density to find a particle at time $t$ at position ${\bf x}$
and ${\bf w}({\bf x},t)=({\rm Re}f_1,{\rm Im}f_1)=\int {\bf p}(\theta) f({\bf x},\theta,t) {\rm d}\theta$ is the polarization density. Here, the field $w:= |{\bf w}|$ 
measures the degree of alignment and ${\bf w}/w$ provides the average swimming direction. 
To close the hierarchy (\ref{ffpk}) we follow the scheme used in \cite{Bertin2009} involving the assumption that 
deviations from isotropy are not too strong. Specifically, we assume that $f_2$, representing nematic order, follows changes in $f_0,f_1$ adiabatically (i.e. $\dot f_2\approx 0$) and that 
modes of order $f_{k_\geq 3} \approx 0$. After some lengthy algebra, this leads to a closed set of equations of motion for $\rho,{\bf w}$:
\ea
\dot \rho &=& - {\rm Pe} \nabla \cdot {\bf w} 
\nonumber \\
\dot {\bf w} &=& -{\bf w} + \4{B\rho}{2}\nabla c - \4{{\rm Pe}}{2}\nabla \rho  + \4{{\rm Pe}^2}{16}\nabla^2 {\bf w} - \4{B^2 |\nabla c|^2}{8} {\bf w} \nonumber \\
&+&
\4{{\rm Pe} B}{16} \left(3 (\nabla {\bf w})^{\rm T}\cdot \nabla c - (\nabla c \cdot \nabla){\bf w} - 3 (\nabla \cdot {\bf w})\nabla c \right) \label{kineqssm} 
\ee

\subsection{Phoretic production}
\label{JP}
Consider an axisymmetrical, half-coated Janus colloid in a coordinate system where the colloids swimming direction is parallel to the $z$-axis (${\bf p}_i={\bf e}_z$) and its midpoint is at ${\bf r}_i$.
Let the colloid produce chemicals or temperature with an overall rate $k_0$, uniformly on the coated hemisphere, i.e. it features a `reaction'-rate-density 
of $\sigma(\theta)=k_0/(2\pi R_0^2)$ for $\theta \in (0,\pi/2)$ ($\theta \in (\pi/2,\pi)$) and zero
elsewhere, for cap-ahead (cap-behind) swimmers (see Fig.~2 in the main text).
We rewrite the overall production by the considered colloid as given by (3) in the main text as
\1 
{P}_i({\bf r},t):= \int {\rm d}\theta \;{\rm d}\phi\; R_0^2 \sin \theta\; \delta\left({\bf r}-({\bf r}_i-\nu R_0 {\bf n})\right)\sigma(\theta) \label{surfint}
\2
where ${\bf n}=(\cos \phi \sin \theta, \sin \phi \sin \theta, \cos \theta)$ is the unit surface normal and $\nu=1$ ($\nu=-1$) for cap-behind (cap-ahead) swimmers.
Expanding the $\delta$-function around ${\bf r}_i$ yields:
\1
\delta\left({\bf r}-({\bf r}_i-\nu R_0 {\bf n}_i)\right) = \delta({\bf r}-{\bf r}_i) +\nu R_0 {\bf n} \cdot \nabla \delta({\bf r}-{\bf r}_i) + \mathcal{O}\left(R_0 {\bf n}\cdot \nabla \delta({\bf r}-{\bf r}_i)\right)^2
\2
Truncating this expansion beyond the explicitly written terms, plugging the result back into (\ref{surfint}) and performing the surface integral yields
\1 
{P}_i({\bf r},t) \approx k_0 \delta({\bf r}-{\bf r}_i) +\nu k_a {\bf p}_i \cdot \nabla \delta({\bf r}-{\bf r}_i) 
\2
where $k_a=k_0 R_0/2$. 
For $N$ colloids we find an overall production of
\1
{P}({\bf r},t):=\Sum{i=1}{N}{P}_i \approx k_0 \rho({\bf r},t) +\nu k_a \nabla \cdot {\bf w}({\bf r},t) \label{prodsm}
\2
which we can use to write the dynamics of the $c$-field (Eq.~(3) in the main text) as
\1
\dot c = \mathcal{D} \nabla^2 c  + K_0 \rho + \nu \4{K_0}{2}\nabla \cdot {\bf w} - K_d c \label{csm}
\2
Together with (\ref{kineqssm}) this equation provides a closed set of continuum equations, allowing us, below, to understand the onset of structure formation. 
The used truncation in (\ref{prodsm}) should be a good approximation to the exact dynamics if the typical interparticle distance is large compared to the colloidal radius; i.e. it should be reliable in the
regime in which we are mainly interested: at low area fractions and close to the uniform phase.
Replacing the half-coating with a point source at one of the intersection points of the colloids surface and symmetry axis while keeping the overall production rate unchanged, 
leads to the same result but with $k_a=k_0 R_0$. Following the instability criteria derived below and discussed in the main text, the Janus-instability criterion depends linearly on $k_a$. 
Thus, a localized 'reaction' source supports this instability compared to a half-sphere coating, while the delay-induced instability and the Keller-Segel instability only depend on the 
overall reaction rate $k_0$ and do therefore not depend on the coating area (for a given swimming speed).
For later convenience, we introduce the dimensionless number $K_a=k_a/(R_0 D_r)$.
\\For colloids without a net production, that produce uniformly on one hemisphere and consume with the same rate on the 
other hemisphere (respectively with a rate of $k_0$), we find: 
\1
P({\bf r},t) \approx 2 \nu k_a \nabla \cdot {\bf w}({\bf r},t) \label{jpr}
\2

\subsection{Linear Stability Analysis}
Here, we perform a linear stability analysis of the uniform phase $(\rho,{\bf w},c)=(\rho_0,0,(K_0/K_d)\rho_0)$ which is a steady state solution of (\ref{kineqssm},\ref{csm}).
As it is not immediate to understand linear stability of the uniform phase in the most general case for our continuum equations, we proceed as follows: 
(i) We first derive a generalized Keller-Segel model (GKS) applying to autophoretic colloids with either attractive or repulsive chemical or thermal interactions, 
whereas the original Keller-Segel applies \cite{Keller1970,Keller1971} to chemoattraction. Our derivation 
is based 
on the assumption that colloids respond quasi-instantaneously to changes in the chemical field
and provides us with simple instability criteria, here expressed in terms of microscopic parameters, 
both for the attractive (Keller-Segel instability) and repulsive (Janus instability \cite{Liebchen2015sm}) phoretic interactions. 
(ii) To better understand the length scale of patterns emerging from the Janus instability, and to account for 'delay effects'
(non-instantaneous response of the orientation field to changes in the phoretic field) which can lead to an additional instability (delay-induced instability \cite{Liebchen2015sm})
we then generalize our analysis:, we will derive a linear stability criterion which is fully representative of the PBP model if the diffusivity of the phoretic field is large and in absence
decay effects of the phoretic field ($K_d=0$). We finally discuss the robustness of our instabilities against finite $K_d$-effects.

\subsubsection{Generalized Keller-Segel model}
Assuming that ${\bf w}$ follows changes in $c,\rho$ adiabatically ($\dot {\bf w}\rightarrow 0$), focusing on the regime of long and intermediate wavelength ($\nabla^2 {\bf w}\rightarrow 0$) and neglecting all terms 
which do not contribute to the dynamics close to the uniform phase (purely nonlinear terms) 
(\ref{kineqssm},\ref{csm}) reduce to a two variable model: 
\ea
\dot \rho &=& -\4{{\rm Pe}}{2} \left(B\nabla \cdot \rho \nabla c - {\rm Pe} \nabla^2 \rho \right) \nonumber \\
\dot c &=& \mathcal{D} \nabla^2 c + K_0 \rho - K_d c + \4{K_a}{2}\left(B\nabla \cdot \rho \nabla c - {\rm Pe} \nabla^2 \rho \right) \label{gkssm}
\ee
Now linearising these equations around their uniform phase solution $(\rho,c)=\left(\rho_0,(K_0/K_d)c\right)$ (in one spatial dimension), writing the result as a linear matrix-vector-equation and calculating the
eigenvalues of the corresponding matrix, the real parts of which determine the linear stability of (\ref{gkssm}), yields the following result:
\\1. Attractive phoretic interactions ($B>0$) destabilize the uniform phase if
\1 
\4{B \rho_0 K_0}{{\rm Pe} K_d}>1 \label{kssm}
\2
This 'Keller-Segel (KS) instability' criterion for chemoattractive Janus colloids, systematically derived here from a microscopic theory, closely resembles the one we derived earlier from a more phenomenological model \cite{Liebchen2015sm}.
Using our link between autophoresis and phoretic interactions ($B=6\pi s {\rm Pe}\mathcal{D}/K_0$, see main text), the KS-instability takes the following form:
\1 \4{6 s f \mathcal{D}}{K_d} > 1 \2
Thus, it occurs generically in Janus colloids if $\mathcal{D} \gg 1$. More specifically, in physical units the KS-instability for Janus colloids can 
be expressed as $6 f D_c/(k_d R_0^2)>1$, i.e. it becomes effective if the 
timescale a localized peak of the phoretic field would need to cover the mean free area per colloid is smaller than ($1/6$ of) the decay time of this field.
The associated instability band is 
\1 0< q^2 < \4{B \rho_0 K_0-{\rm Pe}K_d}{{\rm Pe}\mathcal{D}} \label{KSband} \2
showing that the KS-instability is a long wavelength instability, triggering the growth of long-wavelength fluctuations at the onset of instability and allowing for shorter ones
further away from onset. (For $\mathcal{D}\rightarrow \infty$ the short wavelength edge of the instability band (\ref{KSband}) scales as $q^2 \approx 6 f$, i.e. the associated length scale
reads in physical units $l=2\pi R_0/q \approx 2\pi R_0/\sqrt{6f}$.)
A complete linear stability analysis involving all three fields ($\rho,w,c$) shows that both the instability criterion (\ref{kssm}) and band (\ref{KSband}) hold true exactly for 
attractive Janus colloids. 
In our simulations the KS-instability leads to dense crystal-like clusters which coarsen and merge in the coarse of the dynamics, resulting at late times, in one large macrocluster (Fig.~1A,B in the main text). 
\\\\
2. Considering repulsive phoretic interactions ($B<0$), we find the following instability criterion
\1 
\4{-B K_a \rho_0}{2\mathcal{D}+{\rm Pe}^2} \nu s >1
\2
The factor $K_a$ reveals that this instability is based on the asymmetry of the production of phoretic fields on the surface of our colloids; for this reason we called it the Janus-instability \cite{Liebchen2015sm}.
Using again $B=6\pi s {\rm Pe}\mathcal{D}/K_0$ and assuming $\mathcal{D}\rightarrow \infty$, this criterion reduces to the generic form
\1 
3 f {\rm Pe} s \4{K_a}{K_0} >1 \rightarrow -\4{3}{2}f {\rm Pe} \nu s >1
\2
where the right inequality applies to Janus colloids with $K_a = \nu K_0/2$
and is relevant for repulsive back producers and for attractive front producers (see Fig.~2 in the main text). 
The GKS predicts an associated instability band of
\1 Q>\4{2 K_d}{-B \rho_0 K_a -{\rm Pe}^2/2 - 2 \mathcal{D}} \2
i.e. the Janus instability is a short-wavelength instability if $K_d >0$. Problematically, however, the instability band does not show a small wavelength 
boundary leading to a divergence in the dispersion relation $\lambda(q \rightarrow \infty)\rightarrow \infty$. 
We will lift this problem by performing a more general linear stability analysis involving all three fields $\rho,{\bf w},c$ in the next paragraph.

\subsubsection{Janus instability and length scale of patterns}
As we have just seen, the Janus instability hinges on the asymmetry of the production of phoretic fields on the colloids surfaces and is independent of the isotropic production (and the decay term). 
To show that the Janus instability can lead to patterns with a well-defined length scale even for $K_d\rightarrow 0$, we now assume $K_d=K_0=0$ which corresponds to Janus colloids producing a 
phoretic field on one hemisphere and degrading or consuming the same field on the other hemisphere. This leads to $k_a=\nu k_0 R_0$ (\ref{jpr}) or, in dimensionless units to $K_a=\nu K_0$
and allows us to perform 
a general linear stability analysis of (\ref{kineqssm}) without requiring further approximations. 
Following a similar procedure as above and using Routh-Hurwitz criteria \cite{Willems1970} to analyse the roots of the characteristic polynomial we find the following instability criterion (in 1D). 
\1 
\4{-B \rho_0 K_a (16 \mathcal{D}+{\rm Pe}^2)}{32 \mathcal{D}^2 + 8 \sqrt{2}\mathcal{D}{\rm Pe}^2 + {\rm Pe}^4}>1 \label{janusfull}
\2
Conversely to what our instability analysis based on the GKS suggests, this instability is generally oscillatory.
Using $B=6\pi s {\rm Pe} \mathcal{D}/K_0$ (main text) and assuming $\mathcal{D}\rightarrow \infty$, this criterion reduces to 
\1 
-3{\rm Pe} f \nu s >1 \label{janussm1}
\2
Note that for pure Janus producers ($K_a=\nu K_0/2$) (\ref{janusfull}) reduces to $-3{\rm Pe} f \nu s>2$ 
which is the same criterion as we obtained from the GKS.

The Janus instability applies to thermophoretic Janus particles and to diffusiophoretic Janus particles with $\mu_r<0$ if they swim cap-behind 
(Fig.~2 in main text), but in addition also to diffusiophoretic swimmers with $\mu_r>0$ (chemoattraction) if they swim cap-ahead.
Remarkably, if and only if $B \rho_0 K_a < 2\mathcal{D}+ {\rm Pe}^2$ (i.e. close enough to the onset of instability) this instability is a short wavelength instability even for $K_d=0$ with an onset wavenumber (defining the length scale
where the instability first emerges) given by
\1
q_0^2 = \4{16 (2\sqrt{2}-1)}{16\mathcal{D}+{\rm Pe}^2} \label{janusl}
\2
Eq.~(\ref{janusl}) predicts that the (onset) length scale $l=2\pi/q_0$ of patterns emerging from the Janus instability increase with the self-propulsion velocity of colloids; the increase is linear for ${\rm Pe}^2\gg \mathcal{D}$
and marginal for $\mathcal{D}\gg {\rm Pe}^2$.

\subsubsection{Delay-induced instability}
We now consider the limiting case where the only the isotropic component (monopole moment) of the production process is relevant ($K_0>0, K_a=0$) and assume again that the lifetime of the 
phoretic field is long ($K_d=0$).
Performing an analogous analysis as above leads us to the following criterion for the oscillatory delay-induced instability 
\1 
\4{-B \rho_0 K_0 {\rm Pe}}{2\mathcal{D}}>1 \rightarrow -3{\rm Pe}^2 f s >1
\2
Here, the delay-induced instability applies to diffusiophoretic Janus particles with $\mu_r<0$ and to thermophoretic Janus particles and yields a long-wavelength instability band, reading for large $\mathcal{D}$
\1 0 < q^2 < 
\4{-B}{2\mathcal{D}^2 {\rm Pe}K_0}
\2
Note that the width of this instability band shrinks with $1/\mathcal{D}$ for large $\mathcal{D}$. 
Despite this shrinking, it turns out to be still an important instability: in our particle based simulations, we observe 
structure formation even for isotropic production for large but finite values of $\mathcal{D}=10^3-10^5$. 
We will see in the next paragraph that a finite $K_d$ suppresses instability at long wavelength and turns the delay-induced instability into a short wavelength instability which leads
to pattern formation. 

With significant effort our analysis can be further generalized to derive a combined criterion for the Janus- and the delay-induced instability: 
Proceeding as before, but assuming only $K_d=0$ ($K_a, K_0 \neq 0$), we find:
\1
3{\rm Pe}f\left(\4{K_a}{K_0} +{\rm Pe}\right)>1 \rightarrow 3{\rm Pe}f\left(\4{\nu}{2} +{\rm Pe}\right)>1
\2
Here, the right hand side of the arrow holds true for Janus colloids with $K_a=K_0 \nu/2$ (or $k_a=k_0 R_0 \nu/2$ in physical units).

\subsubsection{Decay effects}
While the instability criterion for the KS instability which we derived above is fully general, for the chemorepulsive instability criteria we have so far assumed 
$K_d \ll 1$, meaning that the lifetime of the phoretic fields is much longer than 
$\sim 1/D_r \sim 10-100s$.
We are now exploring the robustness of our predictions in cases where this assumption is not true and decay processes (or screening effects)
such as bulk reactions for diffusiophoretic colloids degrade the phoretic field on comparatively short timescales.
\\We start our analysis by realizing that $K_d$ is the coefficient of a reaction term, whose main effect is to suppress instabilities at large wavelength. Therefore, we
analyse the roots of the characteristic polynomial similar as before (using Routh-Hurwitz criteria again) but truncate the resulting instability condition at order $q^4$, which makes our analysis
feasible.
Assuming once more that $\mathcal{D}$ is large compared to all other parameters but now allowing $K_d$ alongside with $B$ to be of the same order leads to the following modified criterion for the Janus instability
\1 
\4{4B\rho_0 s K_a}{K_d{\rm Pe}^2+8\mathcal{D}}>1 \Rightarrow \4{\4{3}{2}{\rm Pe}f s \nu}{1+\4{K_d {\rm Pe}}{8 \mathcal{D}}}>1 \label{januskd}
\2
For $K_d \rightarrow 0$ this reduces to our previous result (\ref{janussm1}). Remarkably, as long as $K_d {\rm Pe}^2\ll 8 \mathcal{D}$ the Janus instability is hardly affected by decay processes which 
reflects the short-wavelength character of this instability. 
Numerical calculations of the dispersion relation confirm this prediction and show that the Janus instability may apply even for $K_d\gg \mathcal{D}$ but 
then require larger (negative) $\beta$-values than predicted by (\ref{januskd}).

Not surprisingly, the Delay-induced instability, which is typically ($\mathcal{D}\gg 1$) effective at finite but typically small wavenumbers 
is comparatively sensitive to degradation processes. 
\\To quantify this, we perform a similar approach but now assume $K_a=0$ For $\mathcal{D}\gg 1$ we find
\1 
\4{(-B/\mathcal{D})\rho_0 {\rm Pe} K_0}{2+4K_d + 4\sqrt{K_d(1+K_d)}}>1 \rightarrow \4{3f {\rm Pe}^2 s}{4K_d} >1 \label{kdcr}
\2
where we used again $B=6\pi s {\rm Pe} \mathcal{D}/K_0$, and assumed $K_d\gg 1$ to achieve the criterion on the right hand side of the arrow.
Thus, for Janus colloids with ${\rm Pe}=50$ at $f=0.1$ the delay-induced instability should survive at least up to $K_d \sim 20$; that is if the lifetime of the phoretic field amounts at least $1/20$th of the 
rotational diffusion time of the colloids or about $1$ second.
If $K_d$ is not large enough to suppress the delay-induced instability, its main effect is a strong change of the length scale of the emerging patterns.
For $K_d \ll \mathcal{D}$, the long wavelength edge of the Delay-induced instability can be estimated as (in physical units and at leading order in $1/\mathcal{D}$):
\1 
l \sim 2\pi R_0 \sqrt{\4{-3 f {\rm Pe} s [{\rm Pe}+(K_a/K_0)(1+K_d)]}{K_d (1+K_d)}} \label{delle}
\2
That is, for ${\rm Pe}^2\gg K_d$ the characteristic size of waves emerging from the delay-induced instability increases linearly with ${\rm Pe}\propto v$
and decreases with $K_d$.
For $K_d\gg {\rm Pe}^2\gg 1$ it scales as $l \propto \sqrt{{\rm Pe}/K_d}$.

We finally note that in presence of short-range repulsions among colloids, as in our simulations, short wavelength instabilities cannot be effective if the characteristic length scale $l$ is below 
the size of a colloid. Using (\ref{delle}) allows us to calculate that for colloids with ${\rm Pe}=50, f=0.1$
$l$ becomes comparable to $R_0$ for $K_d \sim 10^2-10^3$, which is well above the $K_d$-threshold which suppresses the delay-induced instability. This explains, why it is possible to 
observe the delay-induced instability also in presence of short-range repulsions among colloids.

\subsubsection{Numerical scheme and parameters for movies}
We use standard Brownian particle dynamics with periodic boundary conditions to simulate the many colloid dynamics and couple it to a central-difference, forward Euler finite difference scheme propagating the phoretic field.
Interactions are described using the slightly soft Weeks-Chandler Anderson repulsion among colloids; we use a cell list to 
accelerate simulations. 
We simulate particles as point-producers (and point-consumers in case of the Janus pattern) 
and choose the time steps and the grid spacing underlying the discretization of the chemical field 
small enough that the emerging patterns does not change in any visible 
qualitative sense (we used grid sizes of up to $1000 \times 1000$ for the chemial field to test convergence).
When simulating colloids with consumption on one side we do not allow for consumption of the chemical field towards negative values of course. 
Following the parameter space collapse derived in the main text, 
our phase diagram depends only on 
$f$ and ${\rm Pe}$, but the specific appearance and length scales of the emerging patterns will of course still depend on other parameters. 

Parameters used in simulations and for Fig.~1 of the main text: 
We simulate cap-behind swimmers ($\nu=1$) and $B$ choose according to (7 in the main text). 
Other parameters are as follows:
\\Movie 1 (Keller-Segel instability and collapse): 
$N=10^4$; $f \approx 5\%$ ${\rm Pe}=50; \mathcal{D}=1666$; $K_0=0.26$; $K_d=0.96$; grid for phoretic field $L^p_x\times L^p_y=400\times 400$.
\\Movie 2 (Exploding Clusters):
$N=3\times 10^4; f=14.7\%; {\rm Pe}=100; \mathcal{D}=2667; K_0=0.83; K_d=0.17; L^p_x\times L^p_y=500\times 500$
\\Movie 3 (Continuously moving pattern):
$N=10^5; f=12\%; {\rm Pe}=100; \mathcal{D}=2667; K_0=0.83; K_d=0.17; L^p_x\times L^p_y=450\times 450$.
\\Movie 4 (Travelling bands):
$N=2.5\times 10^4; f=12.3\%; {\rm Pe}=50; \mathcal{D}=1333; K_0=0.42; K_d=0.21; L^p_x\times L^p_y=500\times 80$.
Here we choose a non-quadratic simulation box such that the pattern settles down into a regular stripe pattern on accessible timescales. 
We observe analogous results for smaller systems in quadratic boxes. 
\\Movie 5 (Janus instability):
$N=2800; f\approx 15\%;  {\rm Pe}=100; \mathcal{D}=2000; K_d=0.5; L^p_x\times L^p_y=600\times 600$;
and $K_0=1.25$ for point source on producing hemisphere and $K_0=-1.25$ on the other one.


\begin{thebibliography}{9}
\bibliographystyle{apsrev}

\bibitem{Keller1970}
Keller, E.~F \& Segel, L.~A.
{\em J. Theor. Biol.} {\bf 26}, 399 (1970).

\bibitem{Keller1971}
Keller, E.~F \& Segel, L.~A.
{\em J. Theor. Biol.} {\bf 30}, 225 (1971).

\bibitem{Liebchen2015}
B. Liebchen, D. Marenduzzo, I. Pagonabarraga, M.E. Cates, {\em Phys. Rev. Lett.}, {\bf 115}, 25 (2015).

\bibitem{Saha2014}
Saha, S, Golestanian, R, \& Ramaswamy, S.
{\em Phys. Rev. E} {\bf 89}, 062316 (2014).

\bibitem{Meyer2014}
Meyer, M, Schimansky-Geier, L, \& Romanczuk, P.
{\em Phys. Rev. E} {\bf 89}, 022711 (2014).

\bibitem{Pohl2014}
Pohl, O \& Stark, H.
{\em Phys. Rev. Lett.} {\bf 112}, 238303 (2014).

\bibitem{Liebchen2016}
B. Liebchen, M.E. Cates, D. Marenduzzo, {\em Soft Matter}, {\bf 12}, 7259 (2016).

%

\bibitem{Jiang2010}
Jiang, H.-R. and Yoshinaga, N. and Sano, M., {\em Phys. Rev. Lett.} {\bf 26}, 268302 (2010).



\bibitem{Theurkauff2012}
Theurkauff, I, Cottin-Bizonne, C, Palacci, J, Ybert, C, \& Bocquet, L.
{\em Phys. Rev. Lett.} {\bf 108}, 268303 (2012).

\bibitem{Palacci2013}
Palacci, J, Sacanna, S, Steinberg, A.~P, Pine, D.~J, \& Chaikin, P.~M.
{\em Science} {\bf 339}, 936 (2013).


\bibitem{Buttinoni2013}
Buttinoni, I, Bialk\'{e}, J, K\"{u}mmel, F, L\"{o}wen, H, Bechinger, C, \&
 Speck, T.
{\em Phys. Rev. Lett.} {\bf 110}, 238301 (2013).

\bibitem{Bialke2015}
Bialk{\'e}, J. and Speck, T. and L{\"o}wen, H.,
{\em J. Non-Cryst. Solids} {\bf 407}, 367 (2015). 



\bibitem{Anderson1989}
Anderson, J.L., {\em Ann. Rev. Fluid Mech.} {\bf 21}, 61 (1989).

\bibitem{Golestanian2007}
Golestanian, R, Liverpool, T.~B, \& Ajdari, a.
{\em New J. Phys.} {\bf 9}, 126 (2007).

\bibitem{Bickel2014}
Bickel, T., Zecua, G. \& W{\"u}rger, A., {\em Phys. Rev. E}(R) {\bf 89}, 050303 (2014).

\bibitem{Bickel2013}
Bickel, T., Majee, A. \& W{\"u}rger, A., Phys. Rev. E {\bf 89}, 012301 (2013).


\bibitem{activematterreview1}
Marchetti, M.~C, Joanny, J.~F, Ramaswamy, S, Liverpool, T.~B, Prost, J, Rao, M,
  \& Simha, R.~A.
{\em Rev. Mod. Phys.} {\bf 85}, 1143 (2013).

\bibitem{activematterreview2}
Ramaswamy, S.
{\em Annu. Rev. Cond. Matt. Phys.} {\bf 1}, 323 (2010).

\bibitem{Tailleur2008}
Tailleur, J \& Cates, M.
{\em Phys. Rev. Lett.} {\bf 100}, 218103 (2008).

\bibitem{Cates2015}
Cates, M.~E \& Tailleur, J.
{\em Annu. Rev. Condens. Matter Phys.} {\bf 6}, 219 (2015).

\bibitem{Brown2014}
Brown, A.~T. \& Poon, W. {\em Soft Matter} {\bf 10}, 4016 (2014).


\bibitem{Brown2016}
Brown, A.~T., Poon, W., Holm, C. \& de Graaf, J. 
{\em Soft Matter} {\bf 13}, 1200 (2017).




\bibitem{Romanczuk2012}
Romanczuk, P. and B{\"a}r, M.s and Ebeling, W. and Lindner, B. and Schimansky-Geier, L.
{\em Eur. Phys. J} {\bf 202}, 1 (2012).

\bibitem{Buttinoni2012}
Buttinoni, I. and Volpe, G. and K{\"u}mmel, F. and Volpe, G. and Bechinger, C.
{\em J. Phys. Condens. Matter} {\bf 28}, 284129 (2012).

\bibitem{Scagliarini2016} A. Scagliarini, I. Pagonabarraga, arXiv: 1605.03773 (2016).
\bibitem{Wagner2017} M. Wagner, M. Ripoll, arXiv: 1701.07071.v1 (2017).


\bibitem{SM}
See Supplemental Material [url], which includes the following additional references 

\bibitem{Willems1970} 
Willems, J.L. {\em Stability Theory of Dynamical Systems} (Wiley Interscience Division, New York) (1970).

\bibitem{Dean1996}
D. Dean, {\em J. Phys. A}, {\bf 29} L613 (1996).

\bibitem{Bertin2009}
E. Bertin, M. Droz, G. Gr{\'e}goire, {\em J. Phys. A}, {\bf 42}, 445001 (2009).






%
\end{thebibliography}

\begin{thebibliography}{}
\bibitem{Willems1970} 
Willems, J.L. {\em Stability Theory of Dynamical Systems} (Wiley Interscience Division, New York) (1970).

\bibitem{Liebchen2015sm}
B. Liebchen, D. Marenduzzo, I. Pagonabarraga, M.E. Cates, {\em Phys. Rev. Lett.}, {\bf 115}, 258301 (2015).

\bibitem{Liebchen2016sm}
B. Liebchen, M.E. Cates, D. Marenduzzo, {\em Soft Matter}, {\bf 12}, 7259 (2016).

\bibitem{Dean1996}
D. Dean, {\em J. Phys. A}, {\bf 29} L613 (1996).

\bibitem{Bertin2009}
E. Bertin, M. Droz, G. Gr{\'e}goire, {\em J. Phys. A}, {\bf 42}, 445001 (2009).

\bibitem{Keller1970}
Keller, E.~F \& Segel, L.~A.
{\em J. Theor. Biol.} {\bf 26}, 399 (1970).

\bibitem{Keller1971}
Keller, E.~F \& Segel, L.~A.
{\em J. Theor. Biol.} {\bf 30}, 225 (1971).

\end{thebibliography}
\end{document}